\documentclass[fleqn,twoside,usenatbib]{article}
\usepackage[headings]{espcrc2}

\readRCS
$Id: espcrc2.tex,v 1.2 2004/02/24 11:22:11 spepping Exp $
\usepackage{graphicx}
\usepackage[figuresright]{rotating}
\usepackage[square, comma, sort&compress]{natbib}

\newcommand{\AmS}{{\protect\the\textfont2
  A\kern-.1667em\lower.5ex\hbox{M}\kern-.125emS}}
\newcommand{\be}{\begin{equation}}
\newcommand{\ee}{\end{equation}}
\newcommand{\bea}{\begin{eqnarray}}
\newcommand{\eea}{\end{eqnarray}}

\title{Simulations of structure formation in interacting dark energy cosmologies}

\author{M. Baldi \address[ITP]{Institut f\"{u}r Theoretische Physik, Universit\"{a}t Heidelberg, Philosophenweg 16, D-69120 Heidelberg, Germany.}\address[MPA]{Max-Planck-Institut f\"{u}r Astrophysik, Karl-Schwarzschild Strasse 1, D-85748 Garching, Germany.}\thanks{mbaldi@mpa-garching.mpg.de}}
       
\runtitle{Simulations of structure formation in interacting dark energy cosmologies}
\runauthor{M. Baldi}

\begin{document}

\begin{abstract}

The evidence in favor of a dark energy component dominating the Universe, and driving its presently accelerated expansion, has progressively grown during the last decade of cosmological observations. If this dark energy is given by a dynamic scalar field, it may also have a direct interaction with other matter fields in the Universe, in particular with cold dark matter. Such interaction would imprint new features on the cosmological background evolution as well as on the growth of cosmic structure, like an additional long-range fifth-force between massive particles, or a variation in time of the dark matter particle mass.
We review here the implementation of these new physical effects in the N-body code {\small GADGET-2}, and we discuss the outcomes of a series of high-resolution N-body simulations for a selected family of interacting dark energy models, as already presented in Baldi et al. \cite{Baldi_etal_2008}.
We interestingly find, in contrast with previous claims, that the inner overdensity of dark matter halos decreases in these models with respect to $\Lambda $CDM, and consistently halo concentrations show a progressive reduction for increasing couplings.
Furthermore, the coupling induces a bias in the overdensities of cold dark matter and baryons that determines a decrease of the halo baryon fraction below its cosmological value.
These results go in the direction of alleviating tensions between astrophysical observations and the predictions of the $\Lambda $CDM model on small scales, thereby opening new room for coupled dark energy models as an alternative to the cosmological constant.

\vspace{1pc}
\end{abstract}

\maketitle

\section{INTRODUCTION}

The existence of dark energy (DE) and cold dark matter (CDM) as the two most abundant components
of the Universe keeps acquiring support as new and more detailed observational
data become available. Throughout the last decade, a large number of independent and complementary datasets (e.g. \cite{Percival_etal_2001,wmap5,Riess_etal_1998,
  Perlmutter_etal_1999, SNLS}) have shown that the Universe is almost spatially flat, with a total matter density of at most 24\% of the critical density, and the remaining 76\%
being made of a DE component able to drive an accelerated expansion. 
The nature of this dominant component, however, is still unknown and its understanding constitutes a 
serious theoretical challenge. 
Dynamical DE models based on the evolution of a self-interacting scalar field have been proposed \cite{Wetterich_1988, Ratra_Peebles_1988} as a possible way to overcome the severe fine-tuning problems that affect the (otherwise very successful) cosmological constant.
An interesting idea in the context of dynamical DE models is then given by the recently suggested possibility of a direct interaction between the dark energy scalar field and other matter
species in the Universe \citep{Wetterich_1995, Amendola_2000, Farrar2004,
  Gubser2004, Farrar2007}. 
  Understanding  in detail what effects
this interaction imprints on observable features \cite{Bean:2008ac, Bertolami:2007zm, Matarrese_etal_2003, Wang:2006qw,
  Guo:2007zk, Mainini:2007ft, Lee:2006za} is therefore a crucial step for a direct observational test of these models.
  
In this paper, after briefly recalling the essential features of interacting DE models and discussing their implementation
in numerical N-body codes, we present the results of the first fully self-consistent high-resolution
hydrodynamic simulations of cosmic structure formation for a selected
family of interacting DE cosmologies.

The results presented here are discussed in larger detail and in a more complete fashion in \cite{Baldi_etal_2008}.


\section{INTERACTING DARK ENERGY COSMOLOGIES}

Interacting dark energy cosmologies can be described as a multicomponent system where,
although the conservation of the total stress energy tensor ${T{^\mu }}_{\nu}$ is not violated:
\begin{equation} \label{tensor_conserv_total}
\nabla _{\mu }T^{\mu }_{\nu } = \sum_{\alpha} \nabla_\mu T^{ \mu}_{(\alpha) \nu} = 0 \,,
\end{equation}
the individual stress energy tensor ${T{^\mu }}_{\nu (\alpha)}$ of each species $\alpha$ is in general not
conserved.
If so, its divergence has a source term $Q_{(\alpha)\nu}$ accounting for the coupling between the individual species:
\begin{equation} \label{tensor_conserv_alpha}
 \nabla_\mu T^{ \mu}_{(\alpha) \nu} = Q_{(\alpha) \nu} \,,
\end{equation}
and the conservation of the total stress energy tensor (Eqn.~\ref{tensor_conserv_total}) translates into the constraint
\begin{equation} \label{Q_conserv_total}
 \sum_{\alpha} Q_{(\alpha) \nu} = 0 \,.
\end{equation}
We consider then a system described by the Lagrangian:
\be \label{L_phi} {\cal L} =
-\frac{1}{2}\partial^\mu \phi \partial_\mu \phi - U(\phi) -
m(\phi)\bar{\psi}\psi + {\cal L}_{\rm kin}[\psi] \,, \ee 
where the role of the DE is played by a scalar field $\phi $ with self-interaction potential $U(\phi )$, and where the
mass of matter fields $\psi$ coupled to the DE is also a function of $\phi $. 
The choice
$m(\phi)$ then specifies the source term
$Q_{(\phi) \nu}$ via the expression: \be 
\label{general_source}
Q_{(\phi) \nu} = \frac{\partial
  \ln{m(\phi)}}{\partial \phi} \rho_c \, \partial_\nu \phi \,. \ee
Due to the
constraint (\ref{Q_conserv_total}), if DE couples only to cold dark matter (CDM, hereafter denoted with a subscript $c$) 
one has that $Q_{(c) \nu} = - Q_{(\phi) \nu}$. 
In the following we will assume that the latter condition always holds. 
In particular, in a cosmological system composed by DE, CDM, and baryons, we will always assume the baryons (denoted with a subscript $b$) to have no coupling to the DE.

The zero-component of equation (\ref{tensor_conserv_alpha}) provides the
conservation equations for the energy densities of each species: 
\bea
\label{rho_conserv_eq_phi} \rho_{\phi}' &=& -3 {\cal{H}} \rho_{\phi} (1 + w_\phi) - Q_{(\phi)0} \,\,\,\, , \\
\label{rho_conserv_eq_c} \rho_{c}' &=& -3 {\cal{H}} \rho_{c} + Q_{(\phi)0} \,, \\
\label{rho_conserv_eq_b} \rho_{b}' &=& -3{\cal{H}} \rho_{b}\,, 
\eea 
which will determine the background evolution of the Universe.
Here a prime denotes a derivative with respect to the conformal time $\tau $, defined via $d\tau \equiv dt/a$, and ${{\cal H} = a'/a}$ is 
the conformal Hubble function, while $w_{\phi } \equiv p_{\phi }/\rho _{\phi}$ is the equation 
of state of the DE.

We will focus our attention here on a class of models identified by the choice: \be \label{coupling_const}
m(\phi) = m_0 e^{-\beta _{c}(\phi) \frac{\phi}{M}} \Rightarrow Q_{(\phi)0} = - \frac{\beta _{c}(\phi)}{M} \rho_c \phi' \,,\ee
where $M\equiv 1/\sqrt{8\pi G_{N}}$ and $G_{N}$ is Newton's gravitational constant.

This set of cosmologies has been widely investigated, for the case of a constant coupling function,
with regard to its background and linear perturbation features
\cite{Wetterich_1995, Amendola_2000}, to its effects on
structure formation \cite{Amendola_2004, Pettorino_Baccigalupi_2008}, and
also via a first N-body simulation \cite{Maccio_etal_2004}.

By perturbing up to linear order the full system of equations governing our cosmological framework 
it is possible to derive 
an effective acceleration equation for a CDM test particle at a distance $r$ from another CDM particle of mass $\tilde{M}_{c}$,
which takes the form (for a complete and rather compact derivation see \cite{Baldi_etal_2008}):
\begin{equation}
\label{CQ_euler}
\dot{\vec{v}}_{c} = -H \left(1 - \frac{\beta _{c}(\phi)}{M}
  \frac{\dot{\phi}}{H}\right) \vec{v}_{c} - \vec{\nabla }\frac{\tilde{G}_{c}\tilde{M}_{c}}{r} \,.
\end{equation}

This equation is manifestly different in three aspects from the usual newtonian acceleration equation in an expanding Universe:
first, the velocity-dependent term now contains an additional contribution proportional to the DE-CDM coupling $\beta _{c}(\phi )$;
second, the CDM test particle feels an effective
gravitational constant $\tilde{G}_{c}$ given by \cite{Amendola_2004}:
\begin{equation}
\label{G_eff}
\tilde{G}_{c} \equiv G_{N}\left[ 1 + 2 \beta _{c}^{2}(\phi )\right] \,;
\end{equation}
third, as a consequence of the energy exchange with the DE scalar field (Eqn. \ref{rho_conserv_eq_c}), the mass $\tilde{M}_{c}$ of the CDM particle sourcing the gravitational potential changes in time according to the expression: 
\be
 \label{effective_mass} 
 \tilde{M}_c \equiv M_c e^{-\int{\beta _{c}(\phi)\frac{{\rm d}\phi}{{\rm d}a}{\rm d}a}} \,.
\ee 
In the N-body analysis described in the following, we consider $\beta _{c}(\phi )$
to be constant, so that the effective mass formally reads $\tilde{M}_c
\equiv M_c e^{-\beta _{c}\left(\phi-\phi_0\right)} $.

Eqn.~\ref{CQ_euler} is very important for our discussion since it represents the starting point for the implementation of coupled DE models in an N-body code. 
It is essential to stress here the vectorial nature of Eqn.~\ref{CQ_euler}, which is a key point to understand the effects of the new physics induced by the DE-CDM interaction
and hence needs to be properly taken into account for a correct implementation in N-body algorithms.

\section{THE SIMULATIONS}

The main aim of our investigation is to explore the effects that arise in the process
of cosmic structure formation as a consequence of the interaction between the DE scalar field and the
CDM fluid. Our focus will concentrate on the nonlinear regime of structure formation, 
and to this end we study a set of high-resolution cosmological
N-body simulations performed with the code {\small GADGET-2} \cite{gadget-2},
that we suitably modified for this purpose.
Previous attempts to use cosmological N-body
simulations for different flavours of modified Newtonian gravity have been
discussed, for instance, in
\cite{Maccio_etal_2004,Nusser_Gubser_Peebles_2005,Stabenau_Jain_2006,Kesden_Kamionkowski_2006,Springel2007,Laszlo_Bean_2008,
  Sutter_Ricker_2008, Oyaizu_2008,Keselman_Nusser_Peebles_2009} but to our knowledge \cite{Maccio_etal_2004} is the
only previous work focusing on the properties of nonlinear structures in models
of coupled quintessence. 
For this reason we apply our methodology to a series of interacting DE models identical
to the ones considered in \cite{Maccio_etal_2004}, where the only difference with respect to this previous work amounts to updating the cosmological parameters to the latest WMAP \cite{wmap5} results ($\Omega _{c} = 0.213$, $\Omega _{b} = 0.044$, $\Omega _{DE} = 0.743$, $\sigma _{8} = 0.769$, $h = 0.719$, $n = 0.963$).
In these models the scalar field $\phi $ has a Ratra-Peebles
\cite{Ratra_Peebles_1988} self-interaction potential
\begin{equation}
\label{RP_potential}
U(\phi ) = \frac{\Lambda ^{4 + \alpha }}{\phi ^{\alpha }} \,,
\end{equation}
with a slope $\alpha = 0.143$, and with a constant coupling $\beta _{c}$ to CDM particles only, as described above; we label them as RP$n$ for values of the coupling $\beta _{c} = n \times 0.04$, with $n$ ranging from $1$ to $5$, in analogy with \cite{Maccio_etal_2004}.

For all the models -- and for a reference $\Lambda $CDM cosmology with the same cosmological parameters -- we run low-resolution simulations in a large box ($L_{\rm{box}} = 320 h^{-1}$ Mpc, $N_{\rm{part}} = 2 \times 128^{3}$) to test the large-scale behaviour of the different cosmologies and the convergence of our implementation; for four of these models ($\Lambda $CDM, RP1, RP2, RP5) we then run high-resolution simulations in a smaller box ($L_{\rm{box}} = 80 h^{-1}$ Mpc, $N_{\rm{part}} = 2 \times 512^{3}$) to investigate the properties of highly nonlinear structures.

\subsection{Methods}

We now describe one by one the main distinctive features of our models and their implementation
in the N-body code \small{GADGET-2}. For a more detailed description of our numerical approach 
see \cite{Baldi_etal_2008}.

{\em Modified expansion rate - } The interaction between DE and CDM modifies the background evolution through the presence of an early DE component
during all the period of structure formation. The effect of such early DE is to modify the expansion history of the Universe.
For the same set of cosmological parameters at $z=0$, different values of the coupling $\beta _{c}$ will then lead to different expansion histories that will have to be separately 
computed and then used for the N-body time integration.

{\em Mass variation - }\label{mass_variation} As discussed above, the coupled species (CDM in our case) feature an effective
particle mass that changes in time according to 
Eqn.~\ref{effective_mass}.  Therefore, the mass of CDM particles within the simulation box will
have to be accordingly corrected at each timestep, while baryon particles will keep a constant mass.

{\em Velocity-dependent acceleration - } As shown in Eqn.~\ref{CQ_euler}, the coupling induces an additional velocity-dependent term in the acceleration equation for CDM particles.
Therefore, an additional term of the form
\begin{equation}
\label{drag-term}
\vec{a}_{v} \equiv \beta _{c}\frac{\dot{\phi }}{M} \vec{v}
\end{equation}
has to be explicitely added to the acceleration of every CDM particle at each timestep.

{\em Fifth-force - } Finally, one of the most important modifications introduced by the DE-CDM
interaction is the presence of a long-range fifth-force effectively represented by
a modified gravitational constant, formally written
as in Eqn.~\ref{G_eff}, for the gravitational interaction of two CDM particles, 
while any interaction involving a baryon particle will be governed by the standard gravitational constant $G_{N}$.  
This dependence of the gravitational strength on the type of the particles involved in the interaction
requires an N-body code to be able to distinguish among different particle
types in the gravitational force calculation.  In {\small GADGET-2}, the
gravitational interaction is computed by means of a TreePM hybrid method
(see \cite{gadget-2} for details about the TreePM algorithm), so that both
the tree and the particle-mesh algorithms have been modified in order to
account for this effectively species-dependent gravity.

\subsection{Tests}

\begin{figure*}[t]
\includegraphics[scale=0.4]{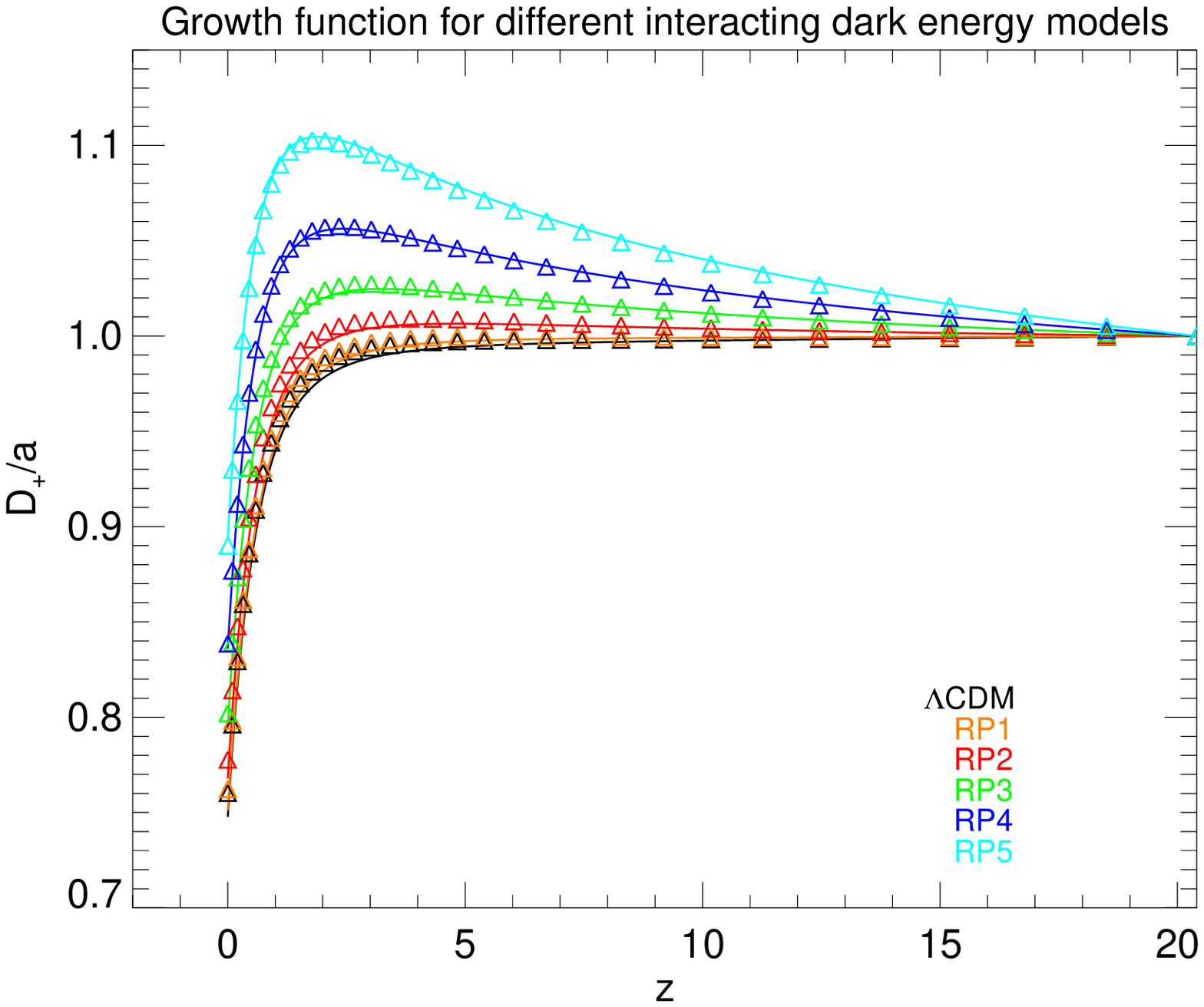}
\includegraphics[scale=0.4]{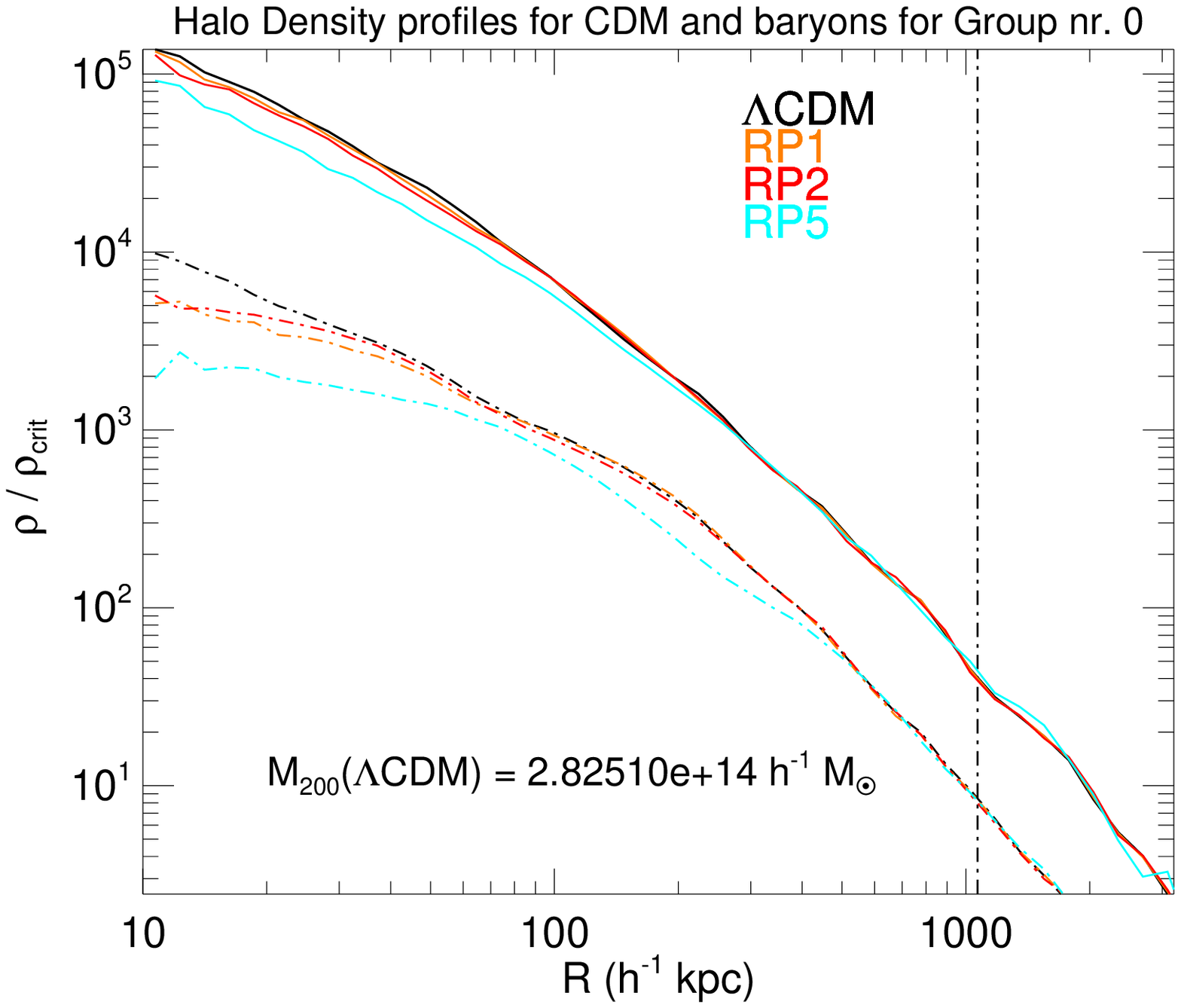}
  \caption{\footnotesize {\em Left Panel:} Evolution of the growth function with redshift for the models
    of coupled DE investigated with the low-resolution simulations. The solid lines are the total
    growth functions as evaluated numerically with {\small CMBEASY}, while the
    triangles are the growth function evaluated from the simulations.
    {\em Right Panel:} Density profiles of CDM (solid lines) and baryons (dot-dashed
    lines) for the most massive halo in the simulation box at
    $z=0$. The vertical dot-dashed line indicates the location of the virial radius for the $\Lambda $CDM halo.}
\label{growth_factor}
\end{figure*}

As a test of our implementation we check whether the linear growth of
density fluctuations in the simulations is in agreement with the linear theory
prediction for each coupled DE model under investigation.  To do so, we
compute the growth factor from the low-resolution simulations and
we compare it with the solution of the system of equations for the evolution of linear
perturbations in interacting DE models, numerically integrated with a suitably modified version of the Boltzmann code {\small
  CMBEASY} \cite{CMBEASY}.  The comparison is shown in the left panel of Fig.~\ref{growth_factor}.
The accuracy of the linear growth computed from the
simulations in fitting the theoretical prediction is of the same order for all
the values of the coupling, and the discrepancy within the two never exceeds
a few percent.

\section{RESULTS}

We present here the most relevant outcomes of the high-resolution simulations of coupled DE
cosmologies described above.

As first basic analysis steps we apply the
Friends-of-Friends (FoF) and {\small SUBFIND} algorithms \cite{Springel2001}
to identify groups and gravitationally bound subgroups in each of our
simulations.  Given that the seed used for the random realization of the power
spectrum in the initial conditions is the same for all the different runs,
structures will form roughly at the same positions in all simulations.
Therefore it is possible, in general, to identify the same objects in all the simulations and to directly compare their properties.
We restrict the comparative analysis of our simulations to the 200 most
massive halos identified by the FoF algorithm, which have virial masses
ranging from $4.64 \times 10^{12} h^{-1} M_{\odot}$ to $2.83 \times 10^{14}
h^{-1} M_{\odot}$.

\subsection{Halo density profiles}

With a suitable procedure we can select, among the 200 most massive halos found at $z=0$ in
each of our simulations, those objects that can be identified with certainty as being the same structure
emerging in the different cosmologies. This leaves 74 objects for our comparison analysis.
For these halos we compute the spherically averaged density profiles of CDM and baryons
as a function of radius around the position of the particle with the minimum
gravitational potential.

Interestingly, the halos formed in the coupled DE cosmologies show
systematically a lower inner overdensity with respect to $\Lambda $CDM, and
this effect grows with increasing coupling. This is clearly visible in the right panel of
Fig.~\ref{growth_factor} where we show the density profiles of CDM and
baryons for the most massive halo in our sample in the four cosmologies with different couplings for which we have high-resolution simulations.
We remark that this result is clearly
incompatible with the essentially opposite behaviour previously reported by
\cite{Maccio_etal_2004}.
\begin{figure*}[t]
\includegraphics[scale=0.4]{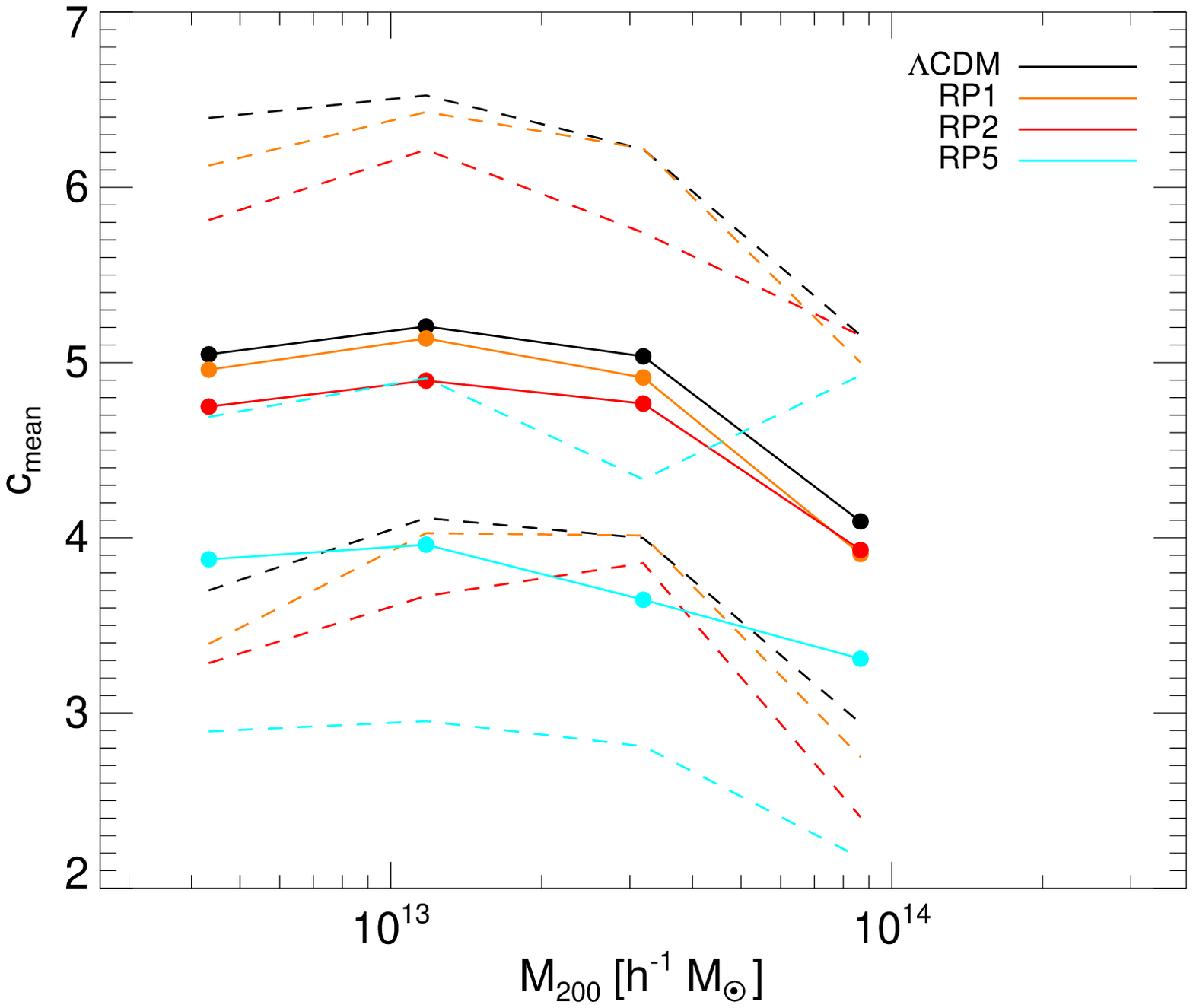}
\includegraphics[scale=0.4]{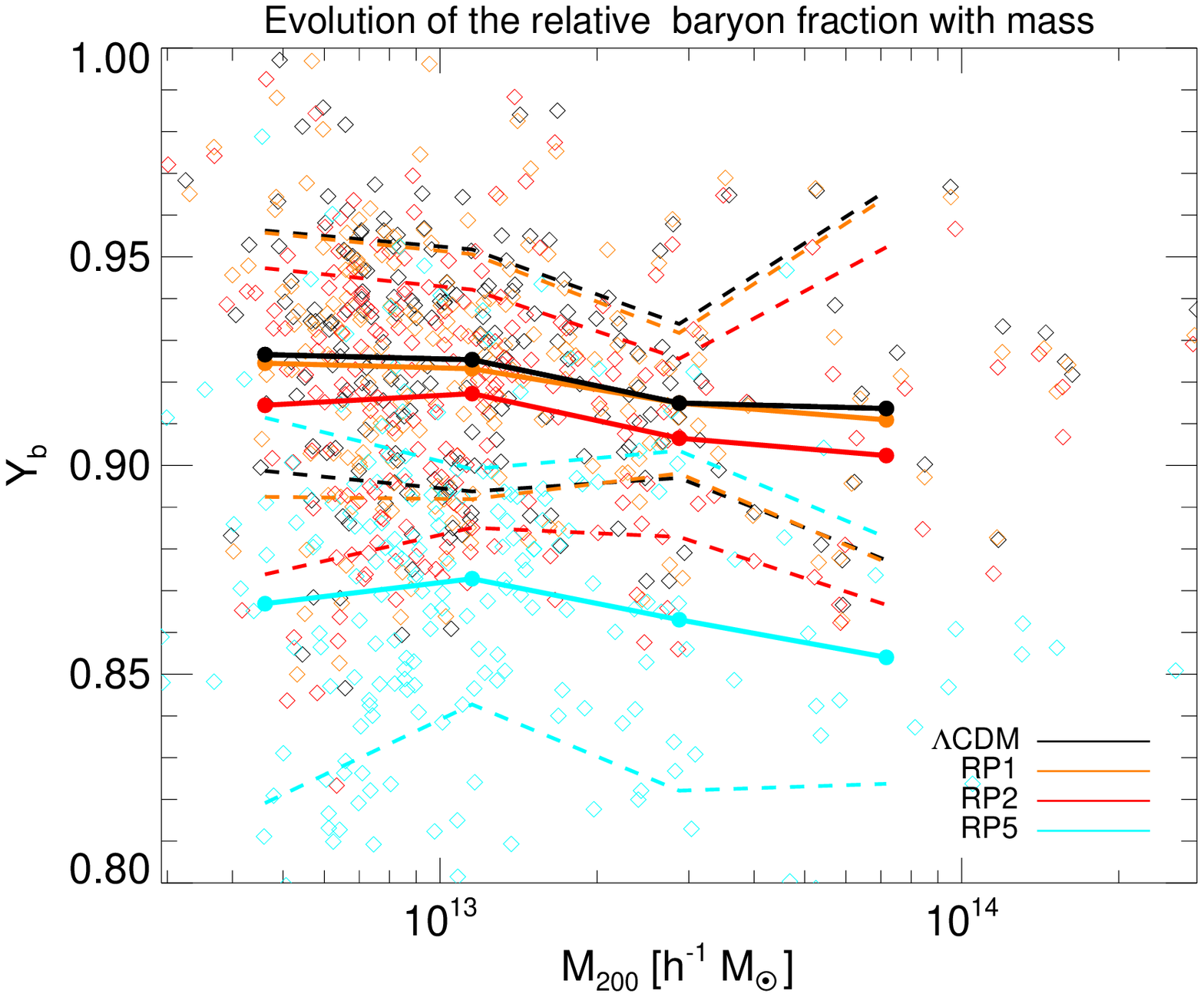}
  \caption{\footnotesize {\em Left Panel:} Evolution of the mean halo concentration as a function of mass for
    the 200 most massive halos in our simulations and for the different
    cosmological models under investigation. 
    The
    halos have been binned by mass, and the mean concentration in each bin is
    plotted as a filled circle. 
    {\em Right Panel:} Evolution with virial mass $M_{200}$ of the relative baryon
    fraction $Y_{b}$ within the virial radius $r_{200}$ for all the halos in
    our sample. The coloured diamonds represent the relative baryon fraction of
    each single halo, while the filled circles and the coloured curves show the
    behaviour of the mean relative baryon fraction in each mass bin for our four
    high-resolution simulations.}
\label{concentrations}
\end{figure*}

In fact, unlike found in \cite{Maccio_etal_2004},
all the halos in our comparison sample 
have density profiles
that are well fit  by the  NFW  fitting function \cite{NFW}
\begin{equation}
\label{NaFrWh}
\left[ \frac{\rho(r) }{\rho _{\rm crit}}\right] _{{\rm NFW}} = \frac{\delta ^{*}}{({r}/{r_{s})}(1+{r}/{r_{s}})^2}\,,
\end{equation}
independent of the value of the coupling.  
The scale radius $r_{s}$ increases for each halo
with increasing coupling $\beta _{c}$, and becomes larger than that found in
$\Lambda $CDM. 
Therefore, the coupling does not affect the overall shape of the density profiles, but rather 
the transition point between the two asymptotic slopes of the NFW function.
In other words, the halos become {\em less concentrated} with
increasing coupling.

\subsection{Halo concentrations}

For all the 200 most massive halos found in each of our 
high-resolution simulations we compute halo concentrations as $c = r_{200}/r_{s}$,
based on our NFW fits to the halo density profiles.  Here $r_{\rm 200}$ is the
radius enclosing a mean overdensity 200 times the critical density. 

Consistently with the trend found for the inner overdensity in the halo
density profiles, we
find that halo concentrations are on average significantly lower for coupled
DE models with respect to $\Lambda $CDM, and the effect again increases with
increasing coupling $\beta_{c}$.  This behaviour is shown explicitly in the
left panel of Fig.~\ref{concentrations}, where we plot halo concentrations as
a function of the halo virial mass $M_{200}$ for our four
high-resolution simulations.

It is possible to show (see \cite{Baldi_etal_2008} for details) that this effect is not due to a change in the formation epoch of structures in the coupled cosmologies with respect to $\Lambda $CDM.
Indeed, a more detailed investigation of this effect shows that the decrease of concentrations arises as a consequence of the {\em expansion} of virialized structures. This expansion is a direct consequence of the DE-CDM interaction, and can be explained as follows: 
on one side, the potential wells of CDM halos become shallower as time goes by as a consequence of the mass decrease of their CDM content, as described by Eqn. \ref{effective_mass};
on the other side, the new velocity-dependent acceleration (Eqn.~\ref{drag-term}) speeds up the motion of CDM particles, thereby injecting energy into the system.
The combination of these two effects determines an overall increase of the total energy of the system, which will then slightly expand to restore virial equilibrium.

It is interesting to note that the effects we find go in the direction of less
``cuspyness'' of halo density profiles, which is preferred by observations and
thus in fact opens up new room for the phenomenology of interacting DE models.

\subsection{Halo baryon fraction}

The extra force proportional to $\beta ^{2}_{c}$ (see Eqn. \ref{G_eff}) that affects only the interaction between two CDM particles 
induces a bias in the evolution of density
fluctuations of baryons and CDM \cite{Mainini:2005fe, Mainini:2006zj, Manera_Mota_2006}.
This bias will appear at all scales during the period of linear growth of density fluctuations, which makes 
it clearly distinguishable from the hydrodynamic bias arising only at small scales as structure evolves.
Furthermore, as it was shown in \cite{Maccio_etal_2004} and \cite{Baldi_etal_2008}, this gravitational bias 
is enhanced in the nonlinear stages of gravitational clustering, and reaches its largest amplitude in the core of
highly overdense regions.

It is interesting that the above effect produces a baryon deficit in
virialized halos, i.e.~they contain fewer baryons than expected based on their
mass and the universal cosmological baryon fraction. In particular, this means
that one can not expect that baryon fractions determined through X-ray
measurements in clusters would yield the cosmological value.  In order to give
a rough estimate of the magnitude of the discrepancy we compute the relative baryon
fraction within the virial radius $r_{200}$ of all the 200 halos in our sample,
defined as
\begin{equation}
Y_{b}\equiv \frac{f_{b}}{\Omega _{b}/\Omega _{m}} \quad $ where$ \quad f_{b} \equiv \frac{M_{b}(<r_{200})}{M_{\rm tot}(<r_{200})}\,.
\label{b_frac}
\end{equation}
For the $\Lambda $CDM case, our results for the evolution of $Y_{b}$ are
consistent with the value of $Y_{b} \sim 0.92$ found by the {\em Santa Barbara
  Cluster Comparison Project} \cite{SBCCP}, and with the more recent results
of \cite{Ettori_etal_2006} and \cite{Gottloeber_Yepes_2007}, while for the
coupled models the relative baryon fraction shows a progressive decrease with
increasing coupling, down to a value of $Y_{b} \sim 0.86-0.87$ for the RP5
case, as shown in the right panel of Fig.~\ref{concentrations}.

It is also important to notice that this effect is always towards lower baryon
fractions in clusters with respect to the cosmological value. This could in
fact alleviate tensions between the high baryon abundance estimated from CMB
observations, and the somewhat lower values inferred from detailed X-ray
observations of galaxy clusters \cite{Vikhlinin_etal_2006,
  McCarthy_etal_2007, LaRoque_etal_2006}.
  
\section{CONCLUSIONS}

We have investigated the effects that arise in the nonlinear regime of structure formation in the context of 
interacting DE models, by means of detailed high-resolution N-body simulations run with a suitably modified version of the
code {\small GADGET-2}.
The numerical implementation we have developed is
quite general and not restricted to the simple specific models of coupled
quintessence that we have discussed in this work. Instead, it should be
well suited for a much wider range of DE models. 

We have presented here the main couclusions of our analysis, concerning the properties of 
collapsed structures in interacting DE models. 
In particular, we have shown that CDM halo density profiles
are remarkably well fit over the resolved range by
the NFW formula for any value of the coupling, but there is a clear trend of a
decrease of the inner halo overdensity with respect to $\Lambda $CDM with
increasing coupling (or, equivalently, an increase of the scale radius $r_{s}$
for increasing coupling). This result conflicts with previous claims for the
same class of coupled DE models \citep{Maccio_etal_2004}.
Consistently, also halo concentrations are reduced with increasing coupling with
respect to $\Lambda $CDM.

Finally, as already shown in \cite{Maccio_etal_2004}, we also find that, as a consequence of
the different effective gravitational strength between CDM and baryons, these two components develop a bias in 
the amplitude of their relative density fluctuations, and this bias is enhanced in the nonlinear region within and
around massive halos. The enhancement
of the bias in highly nonlinear structures has an impact on the determination
of the baryon fraction from cluster measurements, and we have computed for all
our halos its evolution with coupling, finding that the baryon fraction is reduced with increasing
coupling by up to $\sim 8-10\%$ with respect to $\Lambda $CDM for the largest
coupling value we consider.

\section{ACKNOWLEDGMENTS}

I would like to thank the organizers of this {\em Dark Energy Conference}
for the opportunity of presenting these results and for the very nice atmosphere
of the meeting.
I also want to acknowledge V.~Pettorino, G.~Robbers, and V.~Springel for their collaboration
to the work presented in this contribution.
This work has been supported by the TRR33 Transregio Collaborative Research
Network on the ``Dark Universe'', and by the DFG Cluster of Excellence
``Origin and Structure of the Universe''.

\footnotesize
\bibliographystyle{elsevier}
\bibliography{baldi_proceedings_2009_bibliography}

\begin{thebibliography}{10}

\bibitem{Percival_etal_2001}
The 2dFGRS, W.J. Percival et~al.,
\newblock Mon. Not. Roy. Astron. Soc. 327 (2001) 1297, astro-ph/0105252.

\bibitem{wmap5}
WMAP, E. Komatsu et~al.,
\newblock (2008), 0803.0547.

\bibitem{Riess_etal_1998}
Supernova Search Team, A.G. Riess et~al.,
\newblock Astron. J. 116 (1998) 1009, astro-ph/9805201.

\bibitem{Perlmutter_etal_1999}
Supernova Cosmology Project, S. Perlmutter et~al.,
\newblock Astrophys. J. 517 (1999) 565, astro-ph/9812133.

\bibitem{SNLS}
The SNLS, P. Astier et~al.,
\newblock Astron. Astrophys. 447 (2006) 31, astro-ph/0510447.

\bibitem{Wetterich_1988}
C. Wetterich,
\newblock Nucl. Phys. B302 (1988) 668.

\bibitem{Ratra_Peebles_1988}
B. Ratra and P.J.E. Peebles,
\newblock Phys. Rev. D37 (1988) 3406.

\bibitem{Wetterich_1995}
C. Wetterich,
\newblock Astron. Astrophys. 301 (1995) 321, hep-th/9408025.

\bibitem{Amendola_2000}
L. Amendola,
\newblock Phys. Rev. D62 (2000) 043511, astro-ph/9908023.

\bibitem{Farrar2004}
G.R. {Farrar} and P.J.E. {Peebles},
\newblock apj 604 (2004) 1, arXiv:astro-ph/0307316.

\bibitem{Gubser2004}
S.S. {Gubser} and P.J.E. {Peebles},
\newblock prd 70 (2004) 123511, arXiv:hep-th/0407097.

\bibitem{Farrar2007}
G.R. {Farrar} and R.A. {Rosen},
\newblock Physical Review Letters 98 (2007) 171302, arXiv:astro-ph/0610298.

\bibitem{Bean:2008ac}
R. Bean et~al.,
\newblock (2008), 0808.1105.

\bibitem{Bertolami:2007zm}
O. Bertolami, F. Gil~Pedro and M. Le~Delliou,
\newblock Phys. Lett. B654 (2007) 165, astro-ph/0703462.

\bibitem{Matarrese_etal_2003}
S. Matarrese, M. Pietroni and C. Schimd,
\newblock JCAP 0308 (2003) 005, astro-ph/0305224.

\bibitem{Wang:2006qw}
B. Wang et~al.,
\newblock Nucl. Phys. B778 (2007) 69, astro-ph/0607126.

\bibitem{Guo:2007zk}
Z.K. Guo, N. Ohta and S. Tsujikawa,
\newblock Phys. Rev. D76 (2007) 023508, astro-ph/0702015.

\bibitem{Mainini:2007ft}
R. Mainini and S. Bonometto,
\newblock JCAP 0706 (2007) 020, astro-ph/0703303.

\bibitem{Lee:2006za}
S. Lee, G.C. Liu and K.W. Ng,
\newblock Phys. Rev. D73 (2006) 083516, astro-ph/0601333.

\bibitem{Baldi_etal_2008}
M. Baldi et~al.,
\newblock (2008), 0812.3901.

\bibitem{Amendola_2004}
L. Amendola,
\newblock Phys. Rev. D69 (2004) 103524, astro-ph/0311175.

\bibitem{Pettorino_Baccigalupi_2008}
V. Pettorino and C. Baccigalupi,
\newblock Phys. Rev. D77 (2008) 103003, 0802.1086.

\bibitem{Maccio_etal_2004}
A.V. Macci\`{o} et~al.,
\newblock Phys. Rev. D69 (2004) 123516, astro-ph/0309671.

\bibitem{gadget-2}
V. Springel,
\newblock Mon. Not. Roy. Astron. Soc. 364 (2005) 1105, astro-ph/0505010.

\bibitem{Nusser_Gubser_Peebles_2005}
A. Nusser, S.S. Gubser and P.J.E. Peebles,
\newblock Phys. Rev. D71 (2005) 083505, astro-ph/0412586.

\bibitem{Stabenau_Jain_2006}
H.F. Stabenau and B. Jain,
\newblock Phys. Rev. D74 (2006) 084007, astro-ph/0604038.

\bibitem{Kesden_Kamionkowski_2006}
M. Kesden and M. Kamionkowski,
\newblock Phys. Rev. D74 (2006) 083007, astro-ph/0608095.

\bibitem{Springel2007}
V. {Springel} and G.R. {Farrar},
\newblock \mnras 380 (2007) 911, arXiv:astro-ph/0703232.

\bibitem{Laszlo_Bean_2008}
I. Laszlo and R. Bean,
\newblock Phys. Rev. D77 (2008) 024048, 0709.0307.

\bibitem{Sutter_Ricker_2008}
P.M. Sutter and P.M. Ricker,
\newblock (2008), 0804.4172.

\bibitem{Oyaizu_2008}
H. Oyaizu,
\newblock (2008), 0807.2449.

\bibitem{Keselman_Nusser_Peebles_2009}
J.A. Keselman, A. Nusser and P.J.E. Peebles,
\newblock (2009), 0902.3452.

\bibitem{CMBEASY}
M. Doran,
\newblock JCAP 0510 (2005) 011, astro-ph/0302138.

\bibitem{Springel2001}
V. {Springel} et~al.,
\newblock \mnras 328 (2001) 726, arXiv:astro-ph/0012055.

\bibitem{NFW}
J.F. Navarro, C.S. Frenk and S.D.M. White,
\newblock Astrophys. J. 490 (1997) 493, astro-ph/9611107.

\bibitem{Mainini:2005fe}
R. Mainini,
\newblock Phys. Rev. D72 (2005) 083514, astro-ph/0509318.

\bibitem{Mainini:2006zj}
R. Mainini and S. Bonometto,
\newblock Phys. Rev. D74 (2006) 043504, astro-ph/0605621.

\bibitem{Manera_Mota_2006}
M. Manera and D.F. Mota,
\newblock Mon. Not. Roy. Astron. Soc. 371 (2006) 1373, astro-ph/0504519.

\bibitem{SBCCP}
C.S. {Frenk} et~al.,
\newblock \apj 525 (1999) 554, arXiv:astro-ph/9906160.

\bibitem{Ettori_etal_2006}
S. Ettori et~al.,
\newblock Mon. Not. Roy. Astron. Soc. 365 (2006) 1021, astro-ph/0509024.

\bibitem{Gottloeber_Yepes_2007}
S. Gottloeber and G. Yepes,
\newblock Astrophys. J. 664 (2007) 117, astro-ph/0703164.

\bibitem{Vikhlinin_etal_2006}
A. Vikhlinin et~al.,
\newblock Astrophys. J. 640 (2006) 691, astro-ph/0507092.

\bibitem{McCarthy_etal_2007}
I.G. McCarthy, R.G. Bower and M.L. Balogh,
\newblock Mon. Not. Roy. Astron. Soc. 377 (2007) 1457, astro-ph/0609314.

\bibitem{LaRoque_etal_2006}
S. LaRoque et~al.,
\newblock Astrophys. J. 652 (2006) 917, astro-ph/0604039.

\end{thebibliography}


\end{document}